\documentclass[12pt,notitlepage,a4paper]{article}
\usepackage{amssymb}
\usepackage{delarray,amsmath,bbm}
\usepackage{cite}

\usepackage[dvips]{graphicx}
\usepackage{color, graphicx} % for draft version

\newcommand\fverb{\setbox\pippobox=\hbox\bgroup\verb}
\newcommand\fverbdo{\egroup\medskip\noindent%
                              \fbox{\unhbox\pippobox}\ }
\newcommand\fverbit{\egroup\item[\fbox{\unhbox\pippobox}]}
\newbox\pippobox

\newcommand{\mH}{\mathcal{H}}
\newcommand{\mL}{\mathcal{L}}
\newcommand{\mN}{\mathcal{N}}

\newcommand{\half}{\frac{1}{2}}

\newcommand{\pat}{\partial}

\newcommand{\be}{\begin{equation}}
\newcommand{\ee}{\end{equation}}
\newcommand{\bea}{\begin{eqnarray}}
\newcommand{\eea}{\end{eqnarray}}

\newcommand\Dbar{{$\overline {\textrm{D}}\ $}}

\newcommand\Done{{$\overline {\textrm{D}}1\ $}}
\newcommand\Dtwo{{$\overline {\textrm{D}}2\ $}}
\newcommand\Deight{{$\overline {\textrm{D}}8\ $}}

\newcommand\Dbara{{$\overline {\textrm{D}}$}}

\newcommand\Donea{{$\overline {\textrm{D}}1$}}
\newcommand\Dtwoa{{$\overline {\textrm{D}}2$}}

\begin{document}

 \leftline{} \rightline{} \vskip 1cm
 \centerline{\large {\bf Inhomogeneous tachyon dynamics and the zipper}} \vspace{2mm}
 \vskip 1cm
 \renewcommand{\thefootnote}{\fnsymbol{footnote}}
 \centerline{{\bf Niko
 Jokela \footnote{najokela@physics.technion.ac.il} and Matthew
 Lippert \footnote{matthewslippert@gmail.com}
  }}
 \vskip .5cm
  \centerline{\it  Department of Physics}
  \centerline{\it Technion, 32000 Haifa, Israel}
   \centerline{\it and}
 \centerline{\it Department of Mathematics and Physics}
 \centerline{\it University of Haifa at Oranim, 36006 Tivon, Israel}

 \setcounter{footnote}{0}
 \renewcommand{\thefootnote}{\arabic{footnote}}

 \begin{abstract}
We study the process of inhomogeneous tachyon condensation in an intersecting D1- and anti-D1-brane system using an effective tachyon DBI action.   By switching to the Hamiltonian formalism, we numerically solve for the dynamical evolution of the system at a small intersection angle.  We find that the decay proceeds indefinitely and resembles the action of two zippers moving away from the intersection point at the speed of light, zipping the branes together and leaving inhomogeneous tachyon matter behind.  We also discuss the range of validity of our analysis and discuss the relation of the D1-anti-D1 description of the system to one in terms of an intersecting D1-D1-brane pair.

 \end{abstract}

% \hfill   \today

\newpage

%\vskip 1cm

\section{Introduction}

%Frequently in string theory one considers systems that involve intersecting D-branes.
%In some of the cases, one is interested in configurations with two D-branes intersecting at a small angle $\theta\simeq 0$. Physics of such a system can be reasonably well described by an effective non-Abelian DBI action,
%as long as one neglects any interactions between the D-branes. While technically challenging, many interesting phenomena do arise once we allow the D-branes to interact, making it an interesting framework of study. For example, two D-branes, initially intersecting at an angle, can reconnect via open string tachyon condensation. This recombination process plays a central role realizing the Higgs mechanism in Standard Model on intersecting D-branes \cite{higgs} and cosmological models of brane inflation \cite{inflation}.

% tachyons are interesting - non-trivial dynamics, fundamental, reveals connections in strings between open/closed, poorly understood etc
The decay of branes via tachyon condensation is one of the few tractable string theory systems with true nontrivial, nonperturbative dynamical evolution.  The process by which branes decay or annihilate is of fundamental relevancy and reveals important connections between open and closed string sectors.   While much recent progress has been made in understanding open string tachyon physics, the system is still not well understood and many open questions remain.

% branes at angles are interesting - simplest inhomogeneous case, used in Higgs/cosmology, toy model for Sakai-Sugimoto, compact space, inhomogeneous poorly understood, time-independent inhomogeneous -> lower dimensional branes as vortices& solitons
In particular, the annihilation of parallel D-branes and \Dbara-branes is sufficiently simple to allow both a limited worldsheet analysis \cite{Bagchi:2008et} as well as an effective description in terms of a single homogeneous tachyon field \cite{Sen:2002nu,Krausplus}.  The inhomogeneous case, however, is both more interesting and more complicated.  For example, chiral symmetry breaking in the Sakai-Sugimoto model is an example of a physically relevant localized tachyon decay \cite{SStachyons,Jokela:2009tk}.  Much of the effort so far has been limited to studying time-independent inhomogeneous soliton solutions and marginal deformations \cite{Sen:2003tm} rather than on the dynamical evolution.

%branes at angles - simplest inhomogeneous case, applications: Higgs, cosmology, SS
Perhaps the simplest inhomogeneous system with a localized tachyon is where two D-branes intersect at an angle $\theta$.  The tachyonic mode is localized at the intersection, and as it condenses, the two branes reconnect.  This recombination process plays a central role in realizing the Higgs mechanism in Standard Model on intersecting D-branes \cite{higgs} and cosmological models of brane inflation \cite{inflation}.  In addition, this type of localized condensation may serve as a toy model for the reconnection transition found, for example, in the Sakai-Sugimoto model, in non-trivial curved backgrounds. 

For any nonzero value of the angle $\theta$, as mentioned above, the open strings stretching between the branes have tachyonic modes in the low-lying mass spectrum. As is well-known from worldsheet calculations, the negative mass squared of the tachyon in the NS (Neveu-Schwarz) sector increase linearly with $|\theta|$ \cite{Berkooz:1996km}. It is important to notice that when the angle $\theta$ becomes large, these are no longer the lowest mass excited states of the system. In particular, when the intersection angle $\theta$ passes through $\frac{\pi}{2}$ (perpendicular branes) a new tower of states becomes increasingly light and constitute the low-lying mass spectrum of the system \cite{Berkooz:1996km,Arfaei:1996rg,watiaki,SheikhJabbari:1997cv,Ohta:1997fr,Kitao:1998vn,HN,Nagaoka:2003ax,EL}.
Indeed, for the most extreme case $\theta \simeq\pi$, it is more natural to view the system as a D-brane and a \Dbara-brane intersecting at an angle  $\varphi\equiv\pi-\theta$.

Ideally, one would like to study the intersecting D-\Dbar system directly on the worldsheet by calculating the string scattering amplitudes in the rolling tachyon background \cite{Sen:2002nu,Larsenplus}. However, this requires turning on both the inhomogeneous tachyon deformation and a transverse scalar at the same time, a task which is currently out of reach (see \cite{Rey:2003xs,Rey:2003zj,Bagchi:2008et} for work toward this direction in different set-ups). We therefore resort to describing the inhomogeneous tachyon condensation from an effective field theory point of view.

For economic reasons we will consider here intersecting D1-branes.  The generalization to other D$p$-branes and multiple angles is straightforward.

In both flat and curved space, there are known constructions for coincident D-\Dbar effective actions \cite{Krausplus,Sen:2003tm} and the consequent work on understanding their homogeneous decay \cite{Sen:2004nf}. However, when D-branes are even slightly off from being parallel to each other, little is known about their decay process in real time. This is mostly because initially the branes meet at an angle at a single point in spacetime making the tachyon condensation process highly inhomogeneous.

% what we do 
%derive action, tachyon mass
To derive an effective action for a D1-\Done system, we will start from a known action for a coincident D2-\Dtwoa-brane pair with appropriate gauge fields and tachyons turned on.  Then, after T-duality, we will be led to an action describing a D1-\Donea-brane pair which is initially intersecting at an angle $\varphi$ in flat spacetime background.  An analysis of the spectrum of small fluctuations around the false vacuum will give a tachyon mass which, by construction, is exact for small $\varphi$.

%numerical solutions, qualitatively correct
Our main objective is to solve for the explicit temporal evolution, one which we will have to approach numerically.  The inhomogeneous decay is described by two coupled two-dimensional fields whose equations of motion are not amenable to analytic methods.  While we can not entirely trust the quantitative results, because we are solving an effective theory and in addition doing so only approximately, our solutions will hopefully correctly capture the qualitative features of the decay.  

%results - zipper, continues indefinitely, tachyon rolling, dust
The branes decay in roughly two steps.  First, starting from the intersection point, the separated D1 and \Done are attracted to each other in a manner that resembles being zipped together with the zipper traveling asymptotically at the speed of light.  Once the branes are parallel, the tachyon begins to roll with constant velocity toward its vacuum, and, just as in the homogenous case, the dynamics of the branes can then be described in terms of a pressureless, non-interacting tachyon matter.

%matches intuition, DD results, but with DDbar pair
Our results qualitatively match the expected evolution of straight intersecting branes reconnecting into curved separating branes, as has been seen in previous analyses in terms of D1 pairs \cite{HN,EL}, except that in this D1-\Done description the separating branes are connected by a parallel D1-\Done pair with a rolling tachyon.  We will argue, using a slightly modified example, that from the open string point of view, this parallel D1-\Done pair is equivalent to the vacuum, and so that the final states of the D1-\Done and D1-D1 systems in fact differ only by a change of variables.

%organization
This paper is organized as follows.  In Section \ref{sec:review} we will review previous work on tachyon condensation in intersecting D-brane systems, and in Section \ref{sec:derivation} we will turn to deriving an effective action for D1-\Donea-branes.  In addition, we will compute the false vacuum modes and the tachyon mass.  Section \ref{sec:setup} contains some steps necessary for numerical evaluation of the dynamics including a Legendre transformation to the Hamiltonian formalism and a discussion of the boundary conditions.  In Section \ref{sec:analysis} we will present and describe the results of the numerical computations.  We will then explain in  Section \ref{sec:symmetric} the relationship between describing the decay as that of two D-branes rather than as a D-\Dbar pair.  Finally, Section \ref{sec:conc} comprises our summary and ideas for possible extensions of this work.

\section{Review of previous analysis}\label{sec:review}

%In this section we will review some previous work which are relevant for the study of dynamical decay of intersecting D-branes. 
Let us first recall some basic facts from the worldsheet. % and then survey where
%the effective field theory methods match with the exact results, and also where they differ. 
For simplicity, let us start with two parallel D1-branes in Type IIB superstring theory.
%The low-energy description of this system is in terms of a non-Abelian super Yang-Mills theory. 
Then consider dialing the intersection angle $\theta$ to some finite value. The open strings which connect the different D1-branes will then get confined about the intersection point because the tensions of the strings tend to minimize. By analyzing the NS sector of the energy spectrum of these localized strings one finds that the mass squared of the modes behaves linearly with the angle\cite{Berkooz:1996km}, 
\be\label{eq:WSmass}
 m^2_{\rm WS} = \left(N-\half\right)\frac{|\theta|}{\pi\alpha'}\quad ,\quad N=0,1,2,\ldots\ .
\ee
In the rest of the paper we shall only focus on positive angles, $\theta > 0$ and $\varphi = \pi - \theta > 0$, and will henceforth drop the absolute value symbols. We will be interested in the lowest, tachyonic excitation $N=0$. Notice that the tachyonic mode exists at all (nonzero) angles $\theta$ and thus signals the instability of the configuration.

The part of the spectrum in (\ref{eq:WSmass}) is reproduced to order ${\cal O}(\theta)$ by the spectrum of fluctuations around the intersecting D1-D1-brane pair background in a non-Abelian Yang-Mills (YM) theory \cite{HN,EL}.
An intrinsic feature of the effective field theory approach is that the mass of the lowest mode behaves as
\be\label{eq:effmass}
 m^2 = -\frac{\tan\left(\frac{\theta}{2}\right)}{\pi\alpha'}  \ ,
\ee
rather than linearly as in (\ref{eq:WSmass}). From (\ref{eq:effmass}) it is evident that the spectra only match at small angles, but the situation becomes increasingly worse at larger angles. This reflects the fact that the YM description is only viable for small angles. The negative mass squared of the tachyon means that the eigenfunction blows up exponentially in time \cite{HN,EL},
\be\label{eq:tachyoneigen}
 T \sim e^{-i\sqrt{m^2}t} e^{-\frac{\tan\left(\frac{\theta}{2}\right)}{\pi\alpha'} x^2} \ .
\ee
From (\ref{eq:tachyoneigen}) we also see that the tachyon fluctuation modes are localized around the intersection point $x=0$, agreeing with the worldsheet. In \cite{HN} it was further argued, that the geometric realization of the tachyon condensation is a D-brane recombination process. This was explicitly shown by diagonalizing the fluctuations through a local gauge transformation of the brane-coordinates. Here, however, we shall postpone discussing this phenomenon to Section \ref{sec:symmetric}.

One can also use the spectrum (\ref{eq:WSmass}) to check the $\alpha'^2 F^4$ and higher order $\alpha'$ corrections in the expansion of the non-Abelian Dirac-Born-Infeld (DBI) action and find agreement \cite{Nagaoka:2003ax,watiaki,EL}. However, equipped even with the full DBI action, thus far one has not been able to reproduce the mass spectrum (\ref{eq:WSmass}) exactly.\footnote{In \cite{watiaki} it was attempted to fix the discrepancy by considering a symmetric trace prescription of \cite{Tseytlin:1997csa} in the non-Abelian DBI action. Though this procedure did fix the behavior for the lowest excitation, the spacing of the mass spectrum turned out to be incorrect.} 

%\subsection{Perturbative Analysis of D-D system}
%
% YM modes {Hashimoto Nagaoka}
% localized modes
% tachyon mass^2 = - (tan \varphi)^2
% exponentially growing T, separating branes

In the current paper, we are mostly interested in the behavior near $\theta \simeq \pi$. Given the shortcomings of matching with the mass spectrum (\ref{eq:WSmass}) it is better to look for alternative routes.  If one dials the intersection angle all the way to $\theta\to\pi$, one finally arrives to a parallel D1-\Done configuration, which has been investigated in numerous articles. We will therefore find it promising to begin with a well-motivated tachyon DBI action for a D1-\Donea-brane pair, which, by construction, will be valid at small angles $\varphi=\pi-\theta\approx 0$, in contrast to the usual effective action for the D1-D1-brane pair. Derivation of the D1-\Done action shall follow below.

\section{Derivation of the D$1$-\Done action}\label{sec:derivation}

We begin by deriving the effective action for the intersecting D1-\Done system. This can be acquired from the action for a coincident D2-\Dtwoa-brane pair with equal and opposite magnetic fields.  We then perform a T-duality along one of the directions longitudinal to the D2 and \Dtwo which yields the desired D1-\Done action.

The action for a pair of coincident D2-\Dtwo is the usual tachyon DBI type action \cite{Sen:2003tm},
\be
 S_{D2} = -\mu_2\int d^3x \ V(|\tau|)\left(e^{-\phi^{(1)}}\sqrt{-\det\left(  {\mathcal A}^{(1)}_{ab}\right)} + {(1\leftrightarrow 2)}\right) \ ,
\ee
where
\bea
 {\mathcal A}_{ab}^{(i)} & = & \eta_{ab}+\pat_a Z^{(i)}\pat_b Z^{(i)}+2\pi\alpha' F^{(i)}_{ab}+\frac{2\pi\alpha'}{2}(D_a \tau (D_b \tau)^*+ (D_a \tau)^* D_b \tau) \\
 F^{(i)}_{ab} & = & \pat_a A^{(i)}_b-\pat_b A^{(i)}_a \\
 D_a \tau & = & \pat_a \tau-i(A_a^{(1)}-A_a^{(2)})\tau \ . 
\eea
By choosing to work in unitary gauge, we can set ${\rm Im} \ \tau =0$ and will denote ${\rm Re} \ \tau = T$.  With a view towards an upcoming T-duality in the $y$-direction, let us choose an ansatz where $T$ and $A_y$ depend on $t$ and $x$ but not $y$.  We further set $A_y^{(1)} = -A_y^{(2)} \equiv A$.  In addition, we fix the dilaton to be constant $e^{\phi^{(i)}}  =  g_s$ and set the other scalars $Z^{(i)}$ and the other components of the gauge field to zero.  The dynamical variables are then:
\bea
 F_{ty}   & = & \pat_t A_y = \dot A \\
 F_{xy}   & = & \pat_x A_y = A' \\
 D_t T       & = & \pat_0T = \dot T \\
 D_x T       & = & \pat_x T = T' \\
 D_y T       & = & -2iA T 
\eea
With this ansatz, D2 and \Dtwo actions are identical and combine into 
\be
 S_{D2} = -2 \frac{\mu_2}{g_s} \int d^3x \ V(T)\sqrt{-\det\left(\eta_{ab}+2\pi\alpha' F_{ab}+2\pi\alpha' D_aT D_bT\right)} \ .
\ee
Evaluating the determinant and integrating over the circle in the $y$-direction of radius $R$ yields
\bea
 S_{D2} & = & -2(2 \pi R)  \frac{\mu_2}{g_s} \int d^2x V(T) \Bigg(\left(1+ 2\pi\alpha' \left(-{\dot T}^2+{T'}^2\right) \right)\left(1+8\pi\alpha' A^2T^2\right) \nonumber \\
        &   & \qquad\qquad\qquad +(2\pi\alpha')^2 \left(-{\dot A}^2+{A'}^2\right)- (2\pi\alpha')^3 \left(\dot AT'-A'\dot T \right)^2\Bigg)^\half \ .
 \eea

We now T-dualize in $y$ to find the action for intersecting D1-\Donea.  Under the T-duality, $R \to \frac{\alpha'}{R}$, $g_s \to \frac{\sqrt{\alpha'}}{R}g_s $, and $A_y \to \frac{1}{2\pi\alpha'} y$.  This yields the action:
\bea
 S_{D1} & = & -2\frac{\mu_1}{g_s}\int d^2x \ V(T) \\
        &   & \times\sqrt{\left(1-2\pi\alpha' \dot T^2+2\pi\alpha' T'^2\right)\left(1+ \frac{4T^2y^2}{2\pi\alpha' }\right) -\dot y^2 + y'^2 - 2\pi\alpha' (y'\dot T-\dot y T')^2} \ ,\nonumber
\eea
where $\mu_1 = 2\pi \sqrt{\alpha'} \mu_2$.  The scalar $y$ represents half the distance between the D1 and \Donea, so the initial angle between the branes is then given by
\be
\tan{\varphi/2} = \frac{y}{x} \ .
\ee
We can rescale the coordinates $t$ and $x$ and the field $y$ by $(t,x,y) \to \sqrt{2 \pi \alpha'} (t,x,y)$, so we have an action in terms of dimensionless quantities:
\be
\label{action}
 S_{D1} = - \mN \int d^2x \ V( T)\sqrt{\left(1-\dot T^2+ T'^2\right)\left(1+ 4 T^2y^2\right) -\dot y^2 + y'^2 - (y'\dot T-\dot y T')^2} \ ,
\ee
where we define the normalization $\mN = 2(2 \pi \alpha') \frac{\mu_1}{g_s} =  \frac{2}{g_s}$.  Motivated by \cite{KN}, we take the tachyon potential to be 
\be
V(T) = \frac{1}{\cosh(\beta T)} \ ,
\ee
where $\beta = \sqrt{\pi}$ for superstrings.  The vacua for this potential are at $T=\pm\infty$, which implies that to reach them would require an infinite amount of time.

%%%%%%%%%%%%%%%%%%%%%%%%%%%%%%%%%%%%%%%%%%%%%%%%%%%%%%%%%%%%%%%%%%%%%%%%%%%%%%%%%%%%

\subsection{EM tensor for tachyon matter}
For later purposes let us record the energy-momentum tensor for the action (\ref{action}).  Let us generalize the above action (\ref{action}) to curved space as follows
\bea
 S_{\rm curved} & = & - \mN \int dtdx\sqrt{-g} \ V(T) \Bigg(1+g^{ab}(\pat_a y\pat_b y+\pat_a T\pat_b T)  \nonumber\\
                &   & +4T^2 y^2(1+g^{ab}\pat_aT\pat_bT)+(g^{ab}g^{cd}-g^{ac}g^{bd})\pat_a y\pat_b y\pat_c T\pat_dT\Bigg)^\half  \ .
\eea
The energy-momentum tensor is then extracted as
\bea
 T_{ab} & = & - \frac{2}{\sqrt{-g}}\frac{\delta S_{\rm curved}}{\delta g^{ab}}\Big|_{g=\eta} \\
        & = & \eta_{ab} \mL + \frac{\mN  V(T)}{\sqrt{\left(1-\dot T^2+ T'^2\right)\left(1+ 4T^2y^2\right) -\dot y^2 + y'^2 - (y'\dot T-\dot y T')^2}}  \nonumber\\
        &   & \times\Big(\pat_a y\pat_b y+\pat_aT\pat_bT+4T^2 y^2\pat_aT\pat_bT+\pat_a y\pat_b y(\pat T)^2 \nonumber \\
        &   & \qquad\qquad\qquad +\pat_aT\pat_bT(\pat y)^2-\pat_{(a}y\pat_{b)}T\pat y\cdot\pat T\Big) \ .
\eea

Explicitly,
\bea
 T_{00} & = &\mN V(T) \  \frac{ 1+y'^2+ T'^2+4T^2 y^2(1+T'^2)}{\sqrt{\left(1-\dot T^2+ T'^2\right)\left(1+ 4T^2y^2\right) -\dot y^2 + y'^2 - (y'\dot T-\dot y T')^2}} \\
 T_{01} & = &\mN V(T)\ \frac{\dot yy'+\dot T T'+4T^2 y^2\dot T T'}{\sqrt{\left(1-\dot T^2+ T'^2\right)\left(1+ 4T^2y^2\right) -\dot y^2 + y'^2 - (y'\dot T-\dot y T')^2}} \\
 T_{11} & = &\mN V(T) \ \frac{-1+\dot y^2+\dot T^2-4T^2 y^2(1-T^2)}{\sqrt{\left(1-\dot T^2+ T'^2\right)\left(1+ 4T^2y^2\right) -\dot y^2 + y'^2 - (y'\dot T-\dot y T')^2}}\ .
\eea

%%%%%%%%%%%%%%%%%%%%%%%%%%%%%%%%%%%%%%%%%%%%%%%%%%%%%%%%%%%%

\subsection{Mode analysis}

As in the previous studies of D-D systems \cite{EL, HN}, we can analyze the spectrum of fluctuations of the action (\ref{action}) around the initial state to identify the tachyonic mode.    We begin at the maximum of the tachyon potential $T=0$ and will also allow for the possibility of small $x$-dependent corrections to the relative position of the D1 and \Donea, so $y(x) = x \tan(\varphi/2) + \delta y(x)$, where $\delta y \ll 1$.  The action (\ref{action}) expanded to second order in $T$ and $\delta y$ is then 
\bea
S & \sim & \mN \int dx \Bigg(-\frac{\dot T^2}{2} + \frac{T'^2}{2}+\half\left(4x^2 \tan(\varphi/2)^2-\beta^2(1+ \tan(\varphi/2)^2)\right) T^2 \nonumber \\
  &      & \qquad\qquad\qquad + \tan(\varphi/2)\delta y -(\dot\delta y)^2 + (\delta y)^2 \Bigg) \ .
\eea
The position fluctuations $\delta y$ are free and decouple, so at this order the position of the branes is uncorrected.

The equation of motion for the tachyon is
\bea
\label{Teom}
-\ddot T + T'' &=& \left(4x^2 \tan(\varphi/2)^2 - \beta^2 (1+ \tan(\varphi/2)^2)\right) T \ .
\eea
Now, decompose $T$ into modes of definite frequency $\Omega_n$ as
\be
T(t,x) =  \sum_{n = 0}^\infty C_n e^{-i \Omega_n t}T_n(x) \ .
\ee
Normalizability of the modes implies a discrete spectrum, and therefore $n$ is a nonnegative integer.  The equation for the tachyon (\ref{Teom}) now becomes just the Schr\"odinger equation for a harmonic oscillator with mass $m$ and frequency $\omega$, where $m\omega = 2 \tan(\varphi/2)$ and $2mE_n = \beta^2 (1+ \tan(\varphi/2)^2) + \Omega_n^2$.  Imposing the boundary condition $T_n(\infty)=0$, the solution for the modes is therefore
\be
T_n(x) = H_n\left(x\sqrt{2 \tan(\varphi/2)}\right) e^{-x^2 \tan(\varphi/2)} \ ,
\ee
where $H_n$ is the Hermite polynomial of order $n$.  These modes are localized, as expected, near the intersection point at $x=0$.

The frequencies are given by
\be
\Omega_n^2 = 4 \tan(\varphi/2) \left(n + \half\right) -  \beta^2 \left(1+ \tan(\varphi/2)^2\right) \ .
\ee
Focusing on the lowest mode $n=0$ and plugging in $\beta^2 = \pi$, we see that $\Omega_0^2 < 0$ for all values of $\varphi$, giving an exponentially-growing tachyonic mode.  For small $\varphi$,
\be
\label{tachyonmass}
\Omega_0^2 \approx \varphi - \pi
\ee
which matches with the worldsheet calculation (\ref{eq:WSmass}) of the tachyon mass.\footnote{Recall that we are measuring dimensionful quantities in units of $2\pi\alpha'$.}  However, as with the D-D calculations in \cite{EL, HN}, for larger angles the mass computed via the effective action increasingly deviates from the exact result.

%%%%%%%%%%%%%%%%%%%%%%%%%%%%%%%%%%%%%%%%%%%%%%%%%%%%%%%%%%%%%%%%%%%%%%%%%%%%%%%%%%%%

\section{Set-up}\label{sec:setup}

Having arrived at the effective action (\ref{action}), we now need to derive the Hamiltonian equations of motion.  In addition, we will also discuss the appropriate boundary conditions, particularly the modifications required for implementing numerical computations.

\subsection{Hamiltonian formalism}
We will perform a Legendre transformation on the action (\ref{action}) and work in the Hamiltonian formalism \cite{Gibbons:2000hf,Sen:2000kd,others}.  Of course, the Lagrange equations of motion derived from the action (\ref{action}) are in principle equivalent, but for implementing a numerical solution, solving four coupled first-order differential equations was found to be easier than two coupled second-order differential equations.  In addition, using the Hamiltonian framework facilitates a description of the tachyon vacuum, $V(T=\pm\infty)=0$.

The canonical momenta are given by
\bea
\Pi_T & = &  \frac{\partial \mL}{\partial \dot T}  = \frac{\mN V(T)\Big\{\dot T \left(1 + 4y^2 T^2\right)+ y'\left(\dot y T' - y'\dot T \right)\Big\}}{\sqrt{\left(1-\dot T^2+T'^2\right)\left(1+ 4T^2y^2\right) -\dot y^2 + y'^2 - (y'\dot T-\dot y T')^2}} \label{PiTdef} \\
\Pi_y & = &  \frac{\partial \mL}{\partial \dot y}  = \frac{\mN V(T)\Big\{\dot y + T' \left(\dot y T' - y'\dot T\right)\Big\}}{\sqrt{\left(1-\dot T^2+ T'^2\right)\left(1+ 4T^2y^2\right) -\dot y^2 + y'^2 - (y'\dot T-\dot y T')^2}}\label{Piydef} \ .
\eea
The Hamiltonian density is then
\bea
\mH & = & \Pi_T \dot T + \Pi_y \dot y - \mathcal L \\
    & = & \Bigg(\Pi_T^2(1+{T'}^2) + \Pi_y^2 \left(1+4y^2T^2 + {y'}^2\right) + 2y' T' \Pi_y\Pi_T \nonumber\\
    &   & \qquad\qquad\qquad + \mN^2 V(T)^2 \left({y'}^2 + (1+{T'}^2)(1+4y^2T^2)\right) \Bigg)^\half
\eea
from which we derive the equations of motion:%NJ I rewrote eq. PiTEOM to fit into screen...pls check!
\bea
\dot T     & = & \frac{\Pi_T\left(1+T'^2\right)+y' T'\Pi_y}{\mH} \label{TEOM} \\
\dot \Pi_T & = & \frac{-4\Pi_y^2 y^2 T-\mN^2 V(T)^2\left\{4\left(1+T'^2\right) y^2 T+\frac{V'}{V}\left(y'^2 +\left(1+T'^2\right)\left(1+4T^2y^2\right)\right)\right\}}{\mH} \label{PiTEOM}\nonumber \\
           &   & +\partial_x \left(\frac{\Pi_T^2 T'+y'\Pi_T\Pi_y+\mN^2 V(T)^2 T' \left(1+4T^2y^2\right)}{\mH} \right) \\
\dot y     & = & \frac{\Pi_y \left(1+4y^2T^2+y'^2\right)+y' T'\Pi_T}{\mH} \label{yEOM} \\
\dot \Pi_y & = & \frac{-4\Pi_T^2 T^2 y - 4 \mN^2 V(T)^2 \left(1+T'^2\right)T^2 y}{\mH}\nonumber\\
           &   & +\partial_x \left(\frac{\Pi_y^2 y'+T'\Pi_T\Pi_y +\mN^2 V(T)^2 y'}{\mH}\right) \label{PiyEOM} \ .
\eea

\subsection{Boundary conditions}

%initial conditions conditions -  in y, T, y dot , Tdot
For describing the dynamical evolution, we begin at $t=0$ with a straight, static D1 and \Done intersecting at an angle $\varphi$.  This translates to $y(0,x) = x \tan(\varphi/2)$ and $\dot y(0,x) = 0$.  In order to initiate the tachyon rolling, we start it at rest, $\dot T(0,x) = 0$, but slightly displaced from the maximum at $T=0$ near $x=0$.  We do not consider a homogenous tachyon, both because the decay is localized near the intersection and, more importantly, as we will describe, for performing the numerical calculations the tachyon must be zero at the spatial cutoff $x_{\rm max}$.  The initial tachyon profile
\be
T(0,x) = T_\epsilon \left(e^{-x^2} - e^{-x_{\rm max}^2} \right)
\ee
is somewhat arbitrary, but we have checked that the subsequent evolution is largely insensitive to its shape.  At the level of small perturbations at least, we can decompose the initial profile into modes, and while the tachyon mode grows exponentially, the massive modes will just oscillate, contributing some small wiggles.   Changing the overall constant $T_\epsilon$ simply adjusts the timescale of the decay, as we will discuss in the next section.

% conversion into pi_y and pi_T
Since we will be solving the Hamiltonian equations of motion, we need to translate our initial conditions for $\dot T$ and $\dot y$ into conditions for $\Pi_T$ and $\Pi_y$.  From (\ref{PiTdef}) and (\ref{Piydef}) we find that $\Pi_T(t=0,x) = 0$ and $\Pi_y(t=0,x) = 0$.

% large x conditions - in T=0 vacuum at \pm xmax
In addition to initial conditions, we need to impose conditions at the spatial boundaries.  Ideally, we would like to require that at spatial infinities the system is in the false vacuum for all times.   However, for numerical calculations, we employ a finite spatial cutoff, $x \in [-x_{\rm max}, x_{\rm max}]$.  At these boundaries we need to fix the fields, and they must satisfy the equations of motion there.  We therefore fix the finite boundary to be in the false vacuum, i.e., $T(t,\pm x_{\rm max}) = 0$ and $y(t,\pm x_{\rm max}) = \pm x_{\rm max}  \tan(\varphi/2)$.  Of course, we must be sure to choose $x_{\rm max}$ large enough that the spatial cutoff does not affect the dynamics.  Since boundary contributions can only propagate at the speed of light, the region outside the future lightcone of the boundary should be insensitive to the cutoff.  We will be focussed on the relevant physics near the intersection point, so if we consider only times $t < x_{\rm max}$ we will be free of spurious boundary effects.  As a check, we have varied $x_{\rm max}$ and shown that for sufficiently large values, our results are independent of the choice.

%%%%%%%%%%%%%%%%%%%%%%%%%%%%%%%%%%%%%%%%%%%%%%%%%%%%%%%%%%%%%%%%%%%%%%%%%%%%%%%%%%%%

\section{Analysis}\label{sec:analysis}

%solved equations numerically
We solve the Hamiltonian equations (\ref{TEOM}), (\ref{PiTEOM}), (\ref{yEOM}), and (\ref{PiyEOM}) using Mathematica.  For the results presented below, we choose the following representative parameters:
\bea
 \mN         & = & 1 \\
 \varphi      & = & \frac{\pi}{12} \\
 x_{\rm max} & = & 30 \\
 T_\epsilon  & = & 10^{-3} \ .
\eea
We find that the numerical integration is only trustworthy up to some value of $t$.  After this point numerical errors begin to grow large; for example, the stress tensor is no longer even approximately conserved.  %check this
For these parameters, we found that the numerics broke down around $t >10$.

\begin{figure}%[!ht]
%\center{\epsfig{file=TsurfPiover12_090422.eps,width=0.5\textwidth}}
\center{\includegraphics[width=0.5\textwidth]{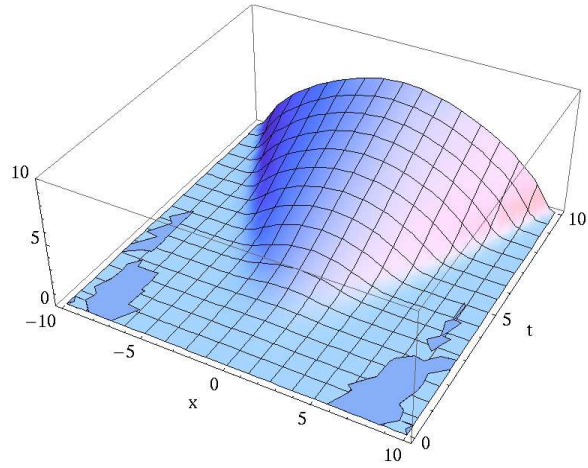}}
\caption{The tachyon as a function of $t$ and $x$.}
\label{fig:Tsurf}
\end{figure}

\begin{figure}%[!ht]
%\center{\epsfig{file=YsurfPiover12_090422.eps,width=0.5\textwidth}}
\center{\includegraphics[width=0.5\textwidth]{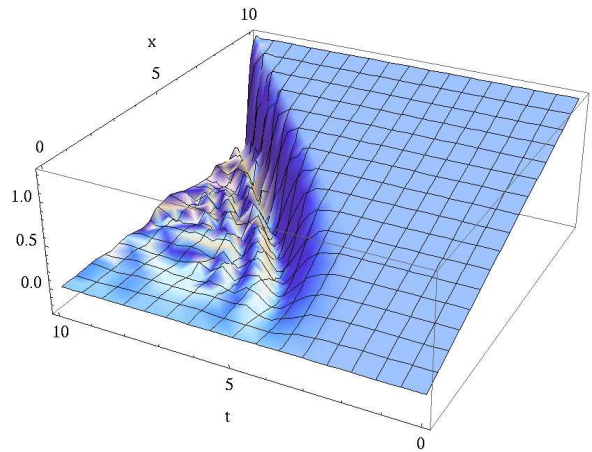}}
\caption{$y$ as a function of $t$ and $x$.}
\label{fig:ysurf}
\end{figure}

%descriptions of T and y surface plots and cross sections - gross features
The solution for the tachyon as a function of $t$ and $x$ is shown in Fig.~\ref{fig:Tsurf}.  Initially, the system remains close to the false vacuum.  But then, near $x=0$, the tachyon begins to grow, and the region of condensation spreads outward roughly at the speed of light.  The corresponding evolution of $y$ is shown in Fig.~\ref{fig:ysurf}.  For clarity, because $y$ is antisymmetric, $y(t,x) = -y(t, -x)$, we only plot positive values of $x$.  Initially, the system again stays near the false vacuum and the branes are straight.  Once the decay process begins, for $x \lesssim t$ the branes are coincident with some small oscillations around $y=0$.

%plots for y and T,  plots of stress tensor

\begin{figure}%[!ht]
\center{\includegraphics[width=0.5\textwidth]{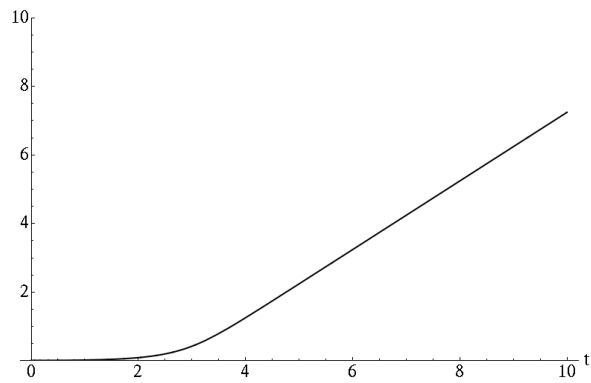}}
\caption{The tachyon at $x=0$ as a function of $t$.}
\label{fig:Tatzero}
\end{figure}

%tachyon at x=0, three regions - false vacuum, transition, rolling
We can more easily illustrate certain features of the evolution by considering cross-sections of Figs.~\ref{fig:Tsurf} and \ref{fig:ysurf}.  In Fig.~\ref{fig:Tatzero} we plot $T(t,0)$.  There are three clearly distinguishable parts of the curve; at early time, for $t \lesssim 3$, the system remains close to the false vacuum $T \approx 0$, then there is a brief transitional region where $T$ begins to condense, rolling down the potential, and then for $t \gtrsim 4$, the tachyon rolls at a constant velocity $\dot T \approx 1$ toward the vacuum at $T=\infty$.

\begin{figure}%[!ht]
\center{\includegraphics[width=0.5\textwidth]{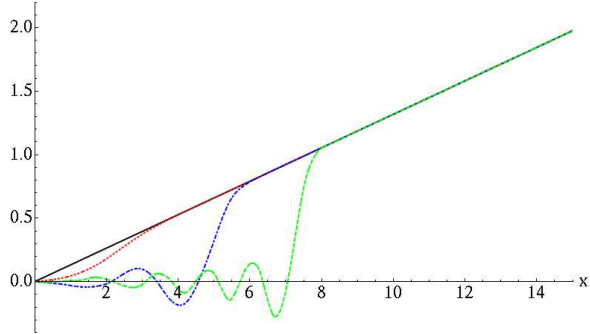}}
\caption{$y$ as a function of $x$, for $t = 2$ (black), 4 (dotted red), 6 (dot-dashed blue), and 8 (dashed green).}
\label{fig:ycross}
\end{figure}

% y profiles, three regions - straight, y->0, wiggles
In addition, we plot in Fig.~\ref{fig:ycross} the profile of $y(x)$ for various fixed times, again only for $x>0$.  For $t=4$, the branes are just beginning to pull toward each other.  At later times, the profile assumes a waterfall shape, again with three regions.  Far from the intersection the branes retain their initial configuration.  But, at around $x \approx t$, a point is reached where $y \to 0$ rapidly, and then, in an increasingly large region around $x=0$, the branes have become coincident except for some small fluctuations.  The process resembles the action of two zippers moving away from the intersection point, zipping the branes together.

\begin{figure}%[!ht]
\center{\includegraphics[width=0.5\textwidth]{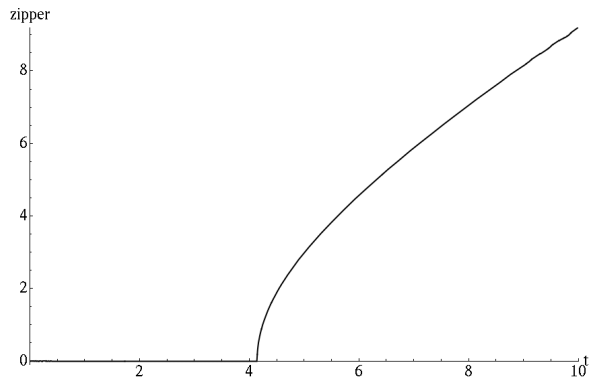}}
\caption{The position of the positive-velocity zipper $z$ as a function of $t$.}
\label{fig:zipper}
\end{figure}

% zipper plot and y -> 0 discussion, wiggles, 
We will define $z(t)$ to be the position of the right-moving zipper and by symmetry the left-moving zipper is at $-z(t)$.  More specifically, for a given time, $z$ equals the largest zero of $y(x)$.  Fig.~\ref{fig:zipper} presents a graph of $z(t)$.  For $t \lesssim 4$, $z=0$ while at late time $\dot z \approx 1$ with a sharp transition in between.

\begin{figure}%[ht]
\center{\includegraphics[width=0.5\textwidth]{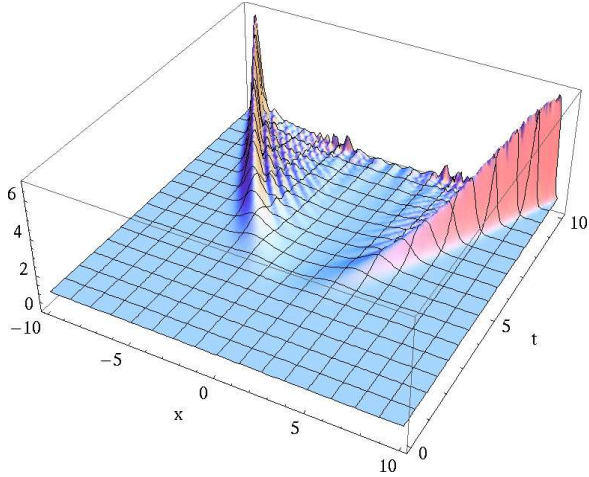}}
\caption{The energy density $T_{00}$ as a function of $t$ and $x$. Notice the large spikes corresponding along the worldlines of the zippers.}
\label{fig:Edensity}
\end{figure}

\begin{figure}%[ht]
\center{\includegraphics[width=0.5\textwidth]{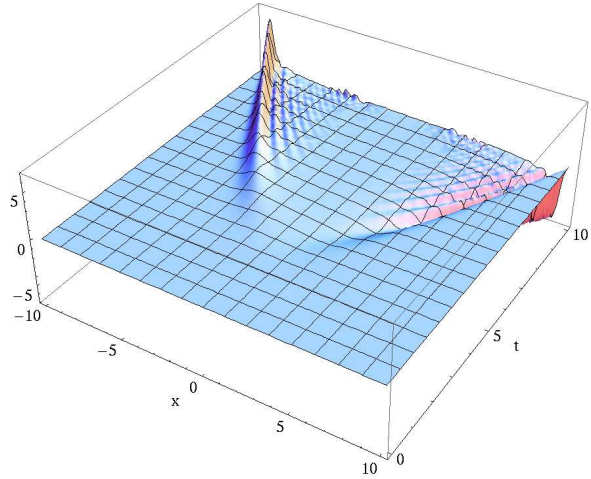}}
\caption{The momentum density $T_{01}$ as a function of $t$ and $x$.  The upward spike corresponds to the worldline of the zipper with negative velocity while the trough corresponds to the zipper with positive velocity.}
\label{fig:Eflow}
\end{figure}

\begin{figure}%[ht]
\center{\includegraphics[width=0.5\textwidth]{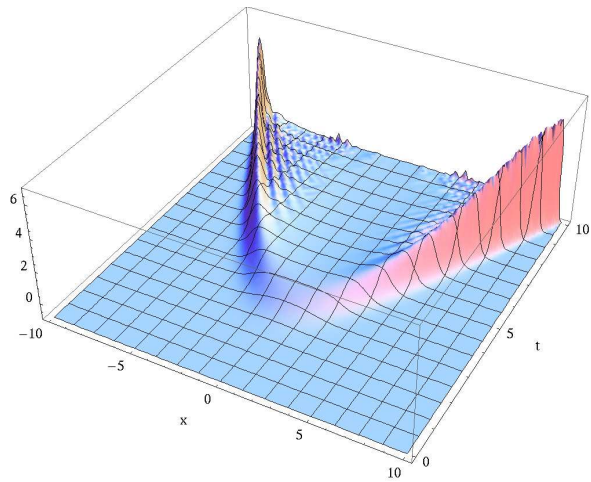}}
\caption{The pressure $T_{11}$ as a function of $t$ and $x$.  Note that the pressure vanishes in the condensing region.}
\label{fig:pressure}
\end{figure}

% stress tensor -  conserved, comparison to BCFT
We can further illustrate the physical properties of the evolution with plots of the components of the stress tensor. Fig.~\ref{fig:Edensity} shows the energy density $T_{00}$.  We have normalized the false vacuum energy so that $T_{00}=1$.  
%, and in the central condensing region $T_{00}$ does something corresponding to the tachyon dust as the system evolves down to the true vacuum.
The motion of the zippers matches roughly with the large spikes in the energy density due to the large kinetic and gradient energy as the branes come together very rapidly.  The position of the zippers is also clearly visible in Fig.~\ref{fig:Eflow} which plots the momentum density $T_{01}$; the zipper moving to the left is represented by the spike and the one going to the right by the trough.  The pressure $T_{11}$ is shown in Fig.~\ref{fig:pressure}.  The false vacuum has negative pressure, and again there are spikes corresponding to the worldlines of the zippers.  Once the tachyon has begun to condense, $T_{11} \to 0$.   In addition, we checked numerically that $T_{\mu\nu}$ was conserved to very good accuracy by the evolution.

% early time timescale tachyon rolling, for zipper to start moving, timescale for y->0 for fixed x
We will first focus on the transitional region between the false vacuum and the rolling region, where the decay first begins.  We can identify three processes which characterize this region, and we can see that they are all related.  Intuitively, the tachyon is massive where the branes are separated by more than the string length, so it cannot really begin rolling until the branes have been zipped together.  However, it is the displacement of the tachyon from zero which gives a potential to $y$ and pulls the branes together.  Consequently, both of these processes must occur together.  The timescale for the onset of the decay is controlled by the initial displacement parameter $T_\epsilon$; the smaller $T_\epsilon$ is, the longer it takes for the decay to begin.  For our choice of $T_\epsilon = 10^{-3}$, the tachyon starts rolling near the intersection at around $t \sim 3$.  At later times $T(t, x)$ starts to grow when $t \sim x$.  Similarly, the zipper begins moving at $t \sim 4$.  As the zipper's velocity approaches the speed of light, a given position on the brane begins to deviate from the false vacuum at around $t \sim x$.

%late time -  description y, T, stress tensor
%tachyon dust -  like parallel case but inhomogeneous
The late time behavior, once the zipper has passed, very much resembles the decay of parallel D1-\Donea-branes but is inhomogeneous.   We actually do not find that $y=0$ past the zipper but rather that $y$ overshoots and oscillates around zero.  As $T$ increases, the masses of these oscillations grow, so their frequencies and amplitudes diminish.  However, we should not really trust our effective action to describe features such as these wiggles which are small compared with the string length.

In the approximation that at late times $y=0$, the dynamics becomes that of parallel branes, with the action for the tachyon becoming just
\be
\label{parallelaction}
 S_{D1} = - \mN \int d^2x \ V(T)\sqrt{1-\dot T^2+ {T'}^2} \ .
\ee
At late time, $T$ is sufficiently large that we may neglect $V(T)$.  The Hamiltonian becomes just
\be
\label{Hamiltoniantachyondust}
\mH = \sqrt{\Pi_T^2(1+{T'}^2)} \ ,
\ee
and the equations of motion  (\ref{TEOM}) and (\ref{PiTEOM}) reduce to
\bea
\label{dusteom}
\dot T  & = & \sqrt{1+{T'}^2} \\
\dot \Pi_T & = &  \partial_x \left(\frac{\Pi_T T'}{ \sqrt{1+{T'}^2}} \right) \ . 
\eea
This is the well-studied pressureless tachyon dust \cite{Gibbons:2000hf,Bergman:2000xf,Sen:2000kd,others,NGothers}
(see also \cite{fluidintep}) where the velocity of the dust is given by
\be
v_\mu = - \partial_\mu T
\ee
and the local rest energy density is 
\be
\epsilon = \frac{\Pi_T}{ \sqrt{1+{T'}^2}} \ .
\ee

\begin{figure}%[ht]
\center{\includegraphics[width=0.5\textwidth]{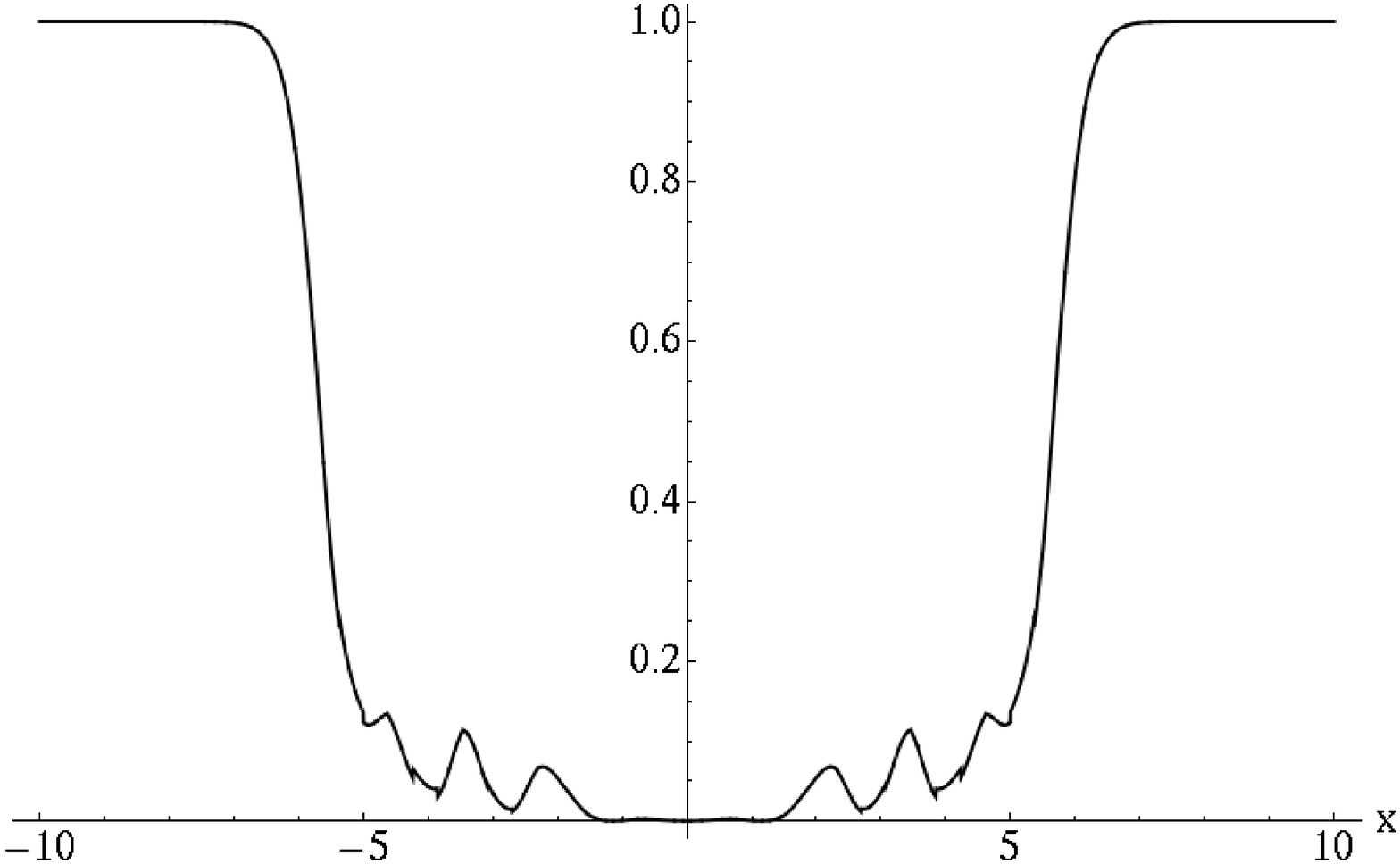}}
\caption{To determine the extent to which the late-time tachyon obeys (\ref{dusteom}), we plot (\ref{dustmeasure}) as a function of $x$ at fixed $t=6$.  This function approximately vanishes in the zipped region, showing that the dust description is valid.}
\label{fig:dusteom}
\end{figure}

We find that the numerical solution accurately reproduces the tachyon dust once the branes have been zipped together.  At $x=0$ where $T'=0$, we saw in Fig.~\ref{fig:Tatzero} that $\dot T \to 1$ in agreement with (\ref{dusteom}).  More generally, we plot in Fig.~\ref{fig:dusteom} the function
\be
\label{dustmeasure}
\frac{1- {\dot T}^2 + {T'}^2}{1+ {\dot T}^2 + {T'}^2}
\ee
which provides a normalized measure of the degree to which (\ref{dusteom}) is satisfied numerically.  To good accuracy, (\ref{dustmeasure}) is zero in the condensing region around $x=0$, implying that in fact (\ref{dusteom}) is approximately obeyed.  Furthermore, we see from Fig.~\ref{fig:pressure} that the condensing region behind the zipper is almost exactly pressureless. 

% interpretation of final state: two separating D1s with parallel annihilating D1-\Done in between corresponding to tachyon matter
% what's up with parallel D1-\Done? Why not vacuum?  See next section for interpretation
If this zipped region is essentially a parallel D1-\Donea, we arrive at the following interpretation for the final configuration of the system.  At late times there are two curved D1-branes separating from each other and straightening but connected to each other by the parallel D1-\Donea.  Based on the results from the D-D description, the two separating D1-branes is expected.  However, rather than being devoid of branes, the region between them contains this decaying D1-\Done pair.  In the next section we will discuss the relation between these two descriptions.

%%%%%%%%%%%%%%%%%%%%%%%%%%%%%%%%%%%%%%%%%%%%%%%%%%%%%%%%%%%%%%%%%%%%%%%%%%%%%%%%%%%%

\section{Relationship between D-D and D-\Dbar systems}\label{sec:symmetric}

% relation between DD and DDbar - change of variables
% D and Dbar labels gauge choice, same DD and D-Dbar same system with theta -> \pi - \varphi, but with different effective descriptions good in different regions
We have been analyzing the system of intersecting D1-branes using a description in terms of a D-\Dbar pair.  This choice of variables, at the level of an exact worldsheet description, amounts to an arbitrary gauge choice; a \Done is simply a D1 with opposite orientation, and a pair of D1-branes intersecting with an angle $\theta$ is equivalent to a D1 crossing a \Done at an angle $\varphi = \pi - \theta$.  However, the two effective field theory descriptions of these systems are inequivalent; each integrates out a different set of massive modes, resulting in different tachyonic and low-mass modes remaining.  For example, using the YM approximation to the non-abelian DBI for the D-D pair, the tachyon's mass (\ref{eq:effmass}) matches the worldsheet value (\ref{eq:WSmass}) only for $\theta \sim 0$, while in the tachyon DBI description of the D-\Dbar system, the tachyon's mass (\ref{tachyonmass}) matches for $\theta \sim \pi$.

% a priori DD and DDbar inequivlanent effective theories, but as we will show qualitative results are the same, though a bit subtle.
Even though the two effective descriptions are not identical and the results need not to match a priori, as we will argue, they actually do produce dynamics which are the same, at least qualitatively.  However, because the change of variables relating the two descriptions is nontrivially complicated, seeing that the results are physically similar can be a bit subtle.  Before considering the relation between the D-D and D-\Dbar systems, we will begin with a simple illustrative example, a case where the two effective descriptions are in fact exactly equivalent and the change of variables between them is explicit.

\begin{figure}%[ht]
\center{\includegraphics[width=0.4\textwidth]{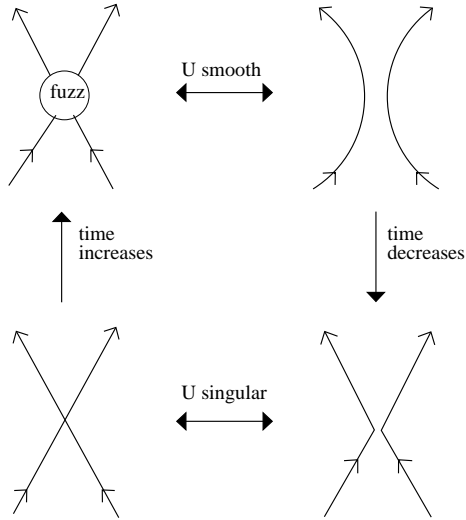}}
\caption{The process of D-D decay and recombination is shown in two gauges.  On the left, the intersection of two straight branes develops off-diagonal elements and becomes fuzzy.  After a gauge transformation $U$, the branes have reconnected and are pulling apart.  However, because $U$ is singular at the beginning of the decay, the initial conditions are inequivalent.}
\label{fig:mixed}
\end{figure}

% decay process description in different gauges: which ends connect to which other ends is gauge dependent 
% DD  fuzz becomes reconnection via  in YM description, cartoon
In the intersecting D-D system, how one describes the decay process depends on one's choice of gauge.  In addition to choosing brane or anti-brane labels, the way the ends of the branes are connected is also to some degree gauge-dependent.  The multiple equivalent descriptions can be easily seen from the non-abelian YM analysis of a D1-D1 decay \cite{HN}, as illustrated in Fig.~\ref{fig:mixed}. 

In the initial variables, the matrix $y$ giving the brane position is diagonal, corresponding to two straight D1-branes.  But, as the decay proceeds, off-diagonal elements begin to grow, fuzzing out the intersection point.  However, $y$ can be rediagonalized by a gauge transformation $U$ which exchanges the brane ends \cite{HN, EL}.  In the new variables, the branes are disconnected, curved, and separating with time.  Evolving back in time to the beginning of the decay, these two bent branes touch at a point.  This odd configuration is, of course, physically inequivalent to the initial straight branes, because at $t=0$ the gauge transformation $U$ relating  them is singular.

The important feature, which also appears in the D-\Dbar picture, is that in the original variables the decaying branes are not seen to annihilate at the intersection point, and the ends remain connected as in the initial state.  Instead, the decay region is characterized by an unusual, non-intuitive state such as the stringy fuzz, and only after a gauge transformation does the system appear as two disconnected branes.

% describe DDbar decay - always have parallel branes in zippered region, gauge equivalent to reconnecting ends with nothing (or closed strings or just no branes) in between
For the description in terms of an intersecting D-\Dbar pair, we showed in Section \ref{sec:analysis} that the branes do not annihilate or reconnect into a disconnected pair of D1-branes.  Such an event is, in fact, outside the range of the tachyon DBI effective theory we used.  Instead, the branes became zipped together, and where the D1 and \Done were parallel, the rolling tachyon was identified as inhomogeneous tachyon matter.   However, although we do not have an explicit gauge transformation analogous to $U$ in the D-D case, there is at least in principle a change of variables such that the parallel D1-\Done are replaced by a gap between a pair of disconnected D1-branes.  Furthermore, in a more complete theory the parallel branes with tachyon matter could be alternatively described by a gas of closed strings and D0-branes (which themselves would decay) between the disconnected D1-branes.

\begin{figure}%[ht]
\center{\includegraphics[width=0.4\textwidth]{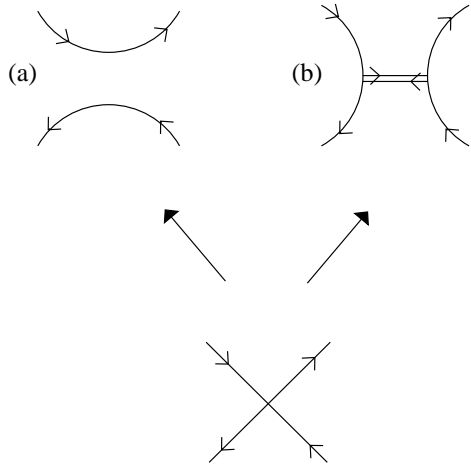}}
\caption{The decay of two angled D1-branes which touch at a point, such that $y(x)$ is symmetric, can proceed in two ways.  Final state (a) consists of a disconnected D1 and \Donea.  Final state (b) also contains a separating D1 and \Donea, but they are connected by a parallel D1-\Donea.}
\label{fig:symmetricdecay}
\end{figure}

% symmetric boundary condition example - different physical system, need to smooth at x=0
Although we can not perform the explicit change of variables for the intersecting D-\Dbar pair, we can illustrate how it works in a similar but physically distinct system.  For the initially straight intersecting D1-\Done considered in Section \ref{sec:analysis}, the relative position of the branes $y$ was always antisymmetric, $y(x) = -y(-x)$.  We can just as easily solve the equations of motion but with $y(x) = y(-x)$ instead, although for numerical computation we have to smooth out the branes at $x=0$ so as to avoid singular first derivatives there.  This symmetric boundary condition corresponds to two angled branes with opposite orientations which touch at the point where they are bent.  In addition, we will choose the intersection angle to be $\varphi = \pi/2$.  This system is unstable, and there are two different directions in which it can decay corresponding to the two ways the branes can recombine, as shown in Fig.~\ref{fig:symmetricdecay}.\footnote{For the numerical computations, which decay mode the system takes depends on how exactly the corner at $x=0$ is resolved.  The details, however, are not that relevant to our discussion.}

\begin{figure}%[!ht]
\center{\includegraphics[width=0.5\textwidth]{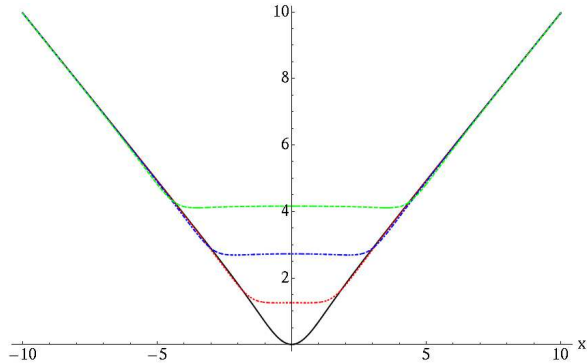}}
\caption{$y$ as a function of $x$, for $t = 0$ (black), 2 (dotted red), 4 (dot-dashed blue), and 6 (dashed green).}
\label{fig:ycrosssymmetric}
\end{figure}

%mode a - like DD
One decay channel, mode (a) in Fig.~\ref{fig:symmetricdecay}, closely resembles the late-time state of intersecting D1-branes, except that instead of ending up with two separating D1-branes, the orientation of one of the D1-branes reversed.   The numerical solution, presented in Fig.~\ref{fig:ycrosssymmetric}, shows the D1 and \Done growing further apart while the tachyon field stays very close to zero. 

%mode b - like D\Dbar
The other decay mode corresponds to the other way of recombining the ends of the branes and is (b) in Fig.~\ref{fig:symmetricdecay}.  In the final state there is still one curved D1 and one curved \Donea, but now they are connected.  Numerically solving the equations of motion yields an evolution qualitatively very similar to those of Section \ref{sec:analysis}; the branes zip together and then the tachyon rolls towards infinity on the parallel D1-\Done pair connecting the two separating D1- and \Donea-branes. 

%both modes are really the same by symmetry, exchange x and y in TDBI
Initially, these two final states appear quite different, but in fact, they must be equivalent due to the symmetry of the initial state which is invariant under the exchange of $x$ and $y$.\footnote{The exchange corresponds to a $\pi/2$ rotation and an orientation reversal.}  This operation should map the final states into each other, and it does up to the parallel D1-\Donea.  However, if we also exchange the roles of $x$ and $y$ in the tachyon DBI action, writing $x$ and $T$ functions of $y$, the parallel D1-\Done would be present in decay (a) and not in mode (b). 

%parallel D-Dbar = vacuum
We conclude that the two modes are, up to this exchange of directions, different descriptions of the same physical state.  In this example the change of variables is particularly simple; whether a parallel D1-\Done connects the separating D1 and \Done or not depends on the choice of the worldvolume coordinates.   At least at the level of open strings where we are working, both the parallel D1-\Done with a rolling tachyon and empty space are equivalent.  In particular, since at late time the parallel D1-\Done with rolling tachyon does not support open string modes, it is effectively as if there are no branes there at all. 

%DD vs DDbar - different approximates of same thing, quantitatively different eg mass; qualitatively the same eg final state
Although the change of variables involved in the intersecting D1-\Done system is much more complicated, the intuition gained from this symmetric example should hold.  Because they are different approximations of the same system, the non-Abelian DBI D-D effective theory and the tachyon DBI D-\Dbar effective theory do not give identical quantitative results.  However,  since we should regard the parallel D1-\Done as essentially the vacuum, both the D-D and D-\Dbar are qualitatively the same.

\section{Summary and discussion}\label{sec:conc}

%what we did - numerical solution for tachyon rolling, inhomogeneous tachyon dust, something about timescales, matches mode analysis  - good for small angle only, 

In this paper we studied the D1-\Done configuration, initially
intersecting at an angle $\varphi$. This configuration is unstable
because tachyonic modes are present at all angles (except for the parallel
D-D-brane pair $\varphi=\pi$). We modeled the evolution of this
system by deriving the equations of motion for the tachyon $T(t,x)$
and the separation field $y(t,x)$ from the tachyon DBI action and
solved them numerically.

We found that, at the very beginning of the evolution, the tachyon,
which was localized at the intersection point, slowly rolled away from
the maximum point of the effective potential. Then, after some time
value, which could be interpreted as the (local) lifetime of the brane
system \cite{GIR}, the tachyon began to grow linearly with time and thus induced
the dynamics for the separation field $y(t,x)$. As time passed, the
branes were pulled toward each other such that the point where they
first met moved from $0$ to larger $\pm x$ values. The process
resembled that of two zippers moving at opposite directions with
speeds of light, zipping the branes together, and continuing indefinitely.  The region between the
zippers had the behavior of a decaying parallel D1-\Done pair, but an
inhomogeneous one. At late time, we were able to capture the essential
features of the remnant by the well-studied inhomogeneous pressureless
tachyon dust.

% compare with other descriptions:  YM, worldsheet; good for small angles, qualitatively the same
Although the tachyon DBI is by construction accurate for $\theta \sim \pi$ while the YM description is valid for $\theta \sim 0$, 
both these results in terms of a D1-\Done pair and the description in terms of intersecting D1-branes give qualitatively the same evolution.  The final state in both pictures contains two reconnected separating D1-branes.  We have argued that the parallel D1-\Done connecting the branes in the tachyon DBI picture is an artifact of the choice of variables and is, in fact, equivalent to the gap between the branes in the YM description.

% could try other DDbar actions especially trace prescriptions (eg Garousi new action), also trace prescription for DD
In both of these effective theories the trace prescriptions for the actions are somewhat ambiguous.  In the case of D-\Dbara, the proposed action of \cite{Garousi:2008nj} gives an alternative way of performing the trace and potentially gives different results from those found here.  The non-Abelian DBI effective description of the D-D system should be valid not just at small angles.  If the correct trace prescription could be found, for example, the worldsheet formula (\ref{eq:WSmass}) for the mass at all angles may be reproduced.

% how well this works as a toy model of chiral symmetry breaking in holographic QCD - qualitatively different, decay doesn't localize/stop
Another potential extension of these results is to other systems featuring localized tachyon condensation.  One may have hoped this simple flat-space system could serve as a toy model for inhomogeneous brane decays in non-trivial curved backgrounds.  In the Sakai-Sugimoto model, for example, a parallel D8-\Deight pair decays into a single U-shaped D8 via the condensation of a tachyon localized at small radii \cite{SStachyons}.  However, one important qualitative difference between this and intersecting branes is that in the flat-space case the decay process continues without end and the condensing region does not stay localized but instead grows without bound.

\bigskip
\noindent

{\bf \large Acknowledgments}

We thank O. Bergman, M. J\"arvinen, R.G. Leigh, G. Lifschytz, and S. Nowling for useful discussions.
This work was supported in part by the Israel Science Foundation under grant no. 568/05.  In addition, N.J. has been supported in part at the Technion by a fellowship from the Lady Davis Foundation.

\end{document}